\documentclass[letterpaper, 10 pt, conference]{ieeeconf}  
\usepackage{graphicx}
\usepackage{algpseudocode}
\usepackage{algorithm}
\usepackage{amsmath} 
\usepackage{xcolor}

\IEEEoverridecommandlockouts                              

\title{\LARGE \bf
CathFlow: Self-Supervised Segmentation of Catheters in Interventional Ultrasound Using Optical Flow and Transformers
}

\author{ Alex Ranne$^{1,*}$, Liming Kuang$^{2,*}$, Yordanka Velikova$^{2,*}$, Nassir Navab$^{2}$, and Ferdinando Rodriguez y Baena$^{1}$
\thanks{*These authors contributed equally to this work}
\thanks{This work was supported by the UKRI CDT in AI for Healthcare under Grant EP/S023283/1, the ICL-TUM Joint Academy of Doctoral Studies (JADS) program, and the TUM Global Incentive Fund.
}
\thanks{$^{1}$Alex Ranne and Ferdinando Rodriguez y Baena are with the Hamlyn Centre for Robotic Surgery, Imperial College London, SW7 2AZ, UK   (e-mail: {\tt\footnotesize \{alex.ranne17, f.rodriguez\}@imperial.ac.uk}) (Corresponding author: Alex Ranne) .}  %
\thanks{$^{2}$Liming Kuang, Yordanka Velikova, and Nassir Navab are with the Chair for Computer Aided Medical Procedures
and Augmented Reality (CAMP), Technical University of Munich, Garching, Germany and also with Munich Center for Machine Learning, Munich, Germany (e-mail: {\tt\footnotesize \{liming.kuang, dani.velikova, nassir.navab, \}@tum.de}) }}

\begin{document}

\maketitle
\thispagestyle{empty}
\pagestyle{empty}

\begin{abstract}

In minimally invasive endovascular procedures, contrast-enhanced angiography remains the most robust imaging technique, but exposes patients and surgeons to prolonged radiation. Alternatives such as ultrasound are difficult to interpret, are highly prone to artifacts and noise, and vary in quality, depending on the experience of the interventional radiologist and machine settings.
In this work, we seek to address both problems by introducing a self-supervised deep learning architecture to segment catheters in longitudinal ultrasound images, without demanding any labeled data. The network architecture builds upon AiAReSeg, a segmentation transformer built with the Attention in Attention mechanism, and is capable of learning feature changes across time and space. To facilitate training, we used synthetic ultrasound data based on physics-driven catheter insertion simulations, and translated the data into a unique CT-Ultrasound common domain, CACTUSS, to improve the segmentation performance. We generated ground truth segmentation masks by computing the optical flow between adjacent frames using FlowNet2, and performed thresholding to obtain a binary mask estimate. Finally, we validated our model on a test dataset, consisting of unseen synthetic data and images collected from silicon aorta phantoms, thus demonstrating its potential for applications to clinical data in the future.

\end{abstract}


\section{INTRODUCTION} \label{Sect: Introduction}

Cardiovascular diseases have been a major source of concern for the WHO, claiming an estimated 17.9 million lives each year \cite{kaplan2016kaplan}. 

Out of these diseases, Abdominal Aortic Aneurysm (AAA) is especially dangerous, identified by the weakening and thinning of the abdominal aorta. In most cases, AAA is asymptomatic, but can lead to severe consequences if ruptured, resulting in a mortality rate of approximately 60\% \cite{ullery2018epidemiology}. 
\begin{figure}[t!]
    \centering
    \includegraphics[width=\columnwidth]{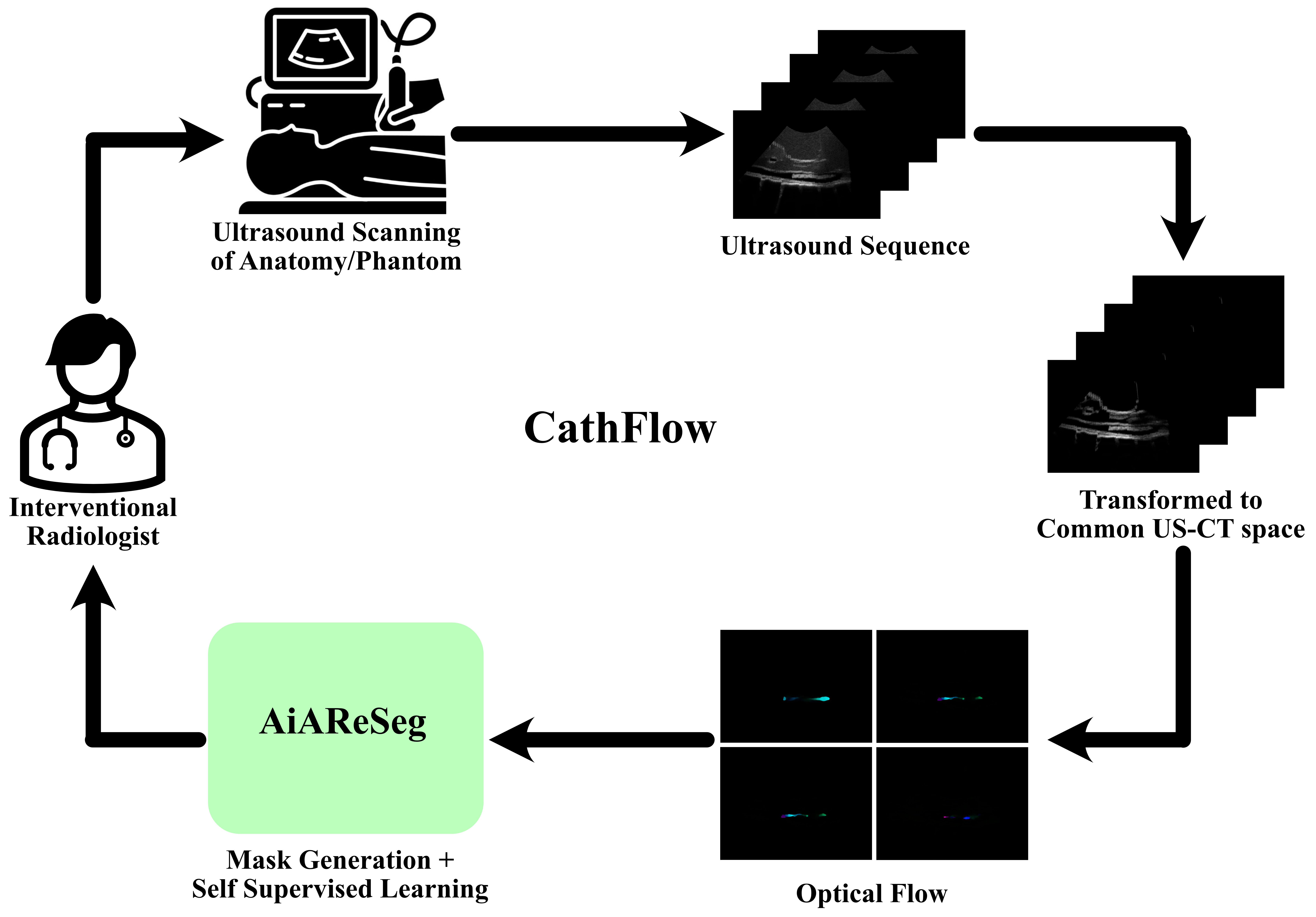}
    \caption{Proposed Interventional Workflow when CathFlow is incorporated into the surgical workflow}
    \vspace{-5mm}
    \label{fig:main}
\end{figure}

In a modern-day operating room, minimally invasive endovascular surgery (MIES) has become the norm for treating such diseases. In MIES procedures, the surgeons start by making a small incision near the patient's groin or the forearm, and a selection of dedicated instruments, such as catheters and guidewires, are manually navigated from the femoral or radial artery up to the diseased lesion under Fluoroscopy. While Fluoroscopy offers the most reliable visual feedback on the position and orientation of the instruments, it emits ionising radiation and is not able to highlight soft tissues. As a result, Digital Subtractive Angiography (DSA), where radiopaque contrast agents are injected into the vasculature, is performed. The toxic nature of the contrast agent, however, constitutes a burden for the patient's kidneys, and may lead to further complications \cite{weisbord2005radiocontrast}.

To mitigate those risks, interventional radiologists seek alternatives, one being intraoperative ultrasound (iUS). Traditionally, contrast-enhanced sonography has been used in endovascular aneurysm repair (EVAR) procedures to localise the landing zone for stent grafts, or for post-operative checkup, ensuring that no endoleaks, is found \cite{kopp2010first}. Similarly, a study has demonstrated the potentials of 3D US in percutaneous transluminal angioplasty (PTA), where US was able to provide sufficient anatomical information, such that it can replace Fluoroscopy in nearly all 4896 cases \cite{wakabayashi2013ultrasound}.

However, employing ultrasound for these procedures has several challenges. Firstly, locating a catheter can be challenging, as it is a slender instrument in a large and complex anatomy. Furthermore, the quality of the image may depend on the experience of the sonographer, the depth of the scan, the force, the orientation of the transducer to ensure good contact, and the nature of the anatomy, just to name a few \cite{gibbs2011ultrasound}. In all cases, US images are prone to visual artifacts and high levels of noise, which makes them difficult to interpret. 

To assist clinicians, computer-assisted systems to extract the vasculature and catheter were developed. While most images are still segmented manually, deep learning-driven image segmentation has been gaining popularity over the years, as it can capture complex underlying behaviours of a scene while learning its physics.
With that said, acquiring a sufficiently large US dataset has proven to be difficult due to a lack of open-source datasets. Labelling such images requires expertise and is time-consuming. In addition, the difficulty in convincing clinicians and regulatory bodies to adopt iUS has also stopped many initiatives in its tracks.

In this paper, we propose a self-supervised transformer framework, trained using synthetic iUS data obtained from a CT to ultrasound translation pipeline. We continue our exploration started from a previous work \cite{ranne2023aiareseg} and provide a solution for improved catheter segmentation in endovascular iUS images, trained from simulation and without labeling. The data is synthesised using CathSim, an open-source physics-based simulator, capable of generating mechanically realistic tissue-instrument interaction (Sect. III-A). Catheter positions during insertion are simulated, then mapped back into CT domain. Consequently, US simulations are generated from the CT labels in the domain of the Common Anatomical CT-US space (CACTUSS) \cite{velikova2024cactuss}. Before training, motion features from adjacent frames inside of a generated US sequence are extracted using FlowNet2, then thresholded and converted to a segmentation mask (Sect. III-B). Finally, using CACTUSS images and estimated ground truth, a transformer-based segmentation network is trained (Sect. III-C), and evaluated on an unseen iUS synthetic dataset, as well as US images collected from a silicon aorta phantom (Sect. IV).

\section{RELATED WORKS}\label{Sect: related works}

\subsection{State of the Art in Image Segmentation}

\subsubsection{Deep Learning: CNN-Based Methods}

In the domain of data-driven methods, Convolutional Neural Networks (CNN) have proven to be a robust technique for automatic medical image segmentation. They demonstrate remarkable performance across different imaging modalities, such as Magnetic Resonance Imaging (MRI) \cite{zaffino2019fully}, Fluoroscopy \cite{nguyen2020end}, Computed Tomography (CT) \cite{litjens2017survey}, and US \cite{jiang2021autonomous, mishra2018ultrasound}. Out of the plethora of CNN-based architectures, U-Net \cite{ronneberger2015u} stands out as one of the most substantial in the field. It features an encoder-decoder architecture, to downsample an image using different CNN kernel sizes until a bottleneck, then upsample it in the decoder and combine with encoder features of varying scales. Further adaptations, such as the nnU-Net \cite{isensee2021nnu}, introduced a preprocessor and hyper-parameter tuning framework, to self-configure to its most appropriate settings. 

\subsubsection{Deep Learning: Transformer and Hybrid Methods}

Following the introduction of the self-attention mechanism \cite{chorowski2015attention}, transformer architectures became pervasive across various domains. Introduced initially by Vaswani et al. \cite{vaswani2017attention} in natural language processing tasks, it quickly emerged as a powerful alternative to CNNs. In the Vision Transformer (ViT) works \cite{dosovitskiy2020vit}, the authors demonstrated its ability to generalise to image patches, while retaining spatial information. Subsequently, Zheng et al. proposed the Segmentation Transformer (SETR) \cite{zheng2021rethinking}, which only uses the attention mechanism together with a Sigmoid activation function to generate segmentation masks.
Combining the benefits of CNN in extracting features at different scales, and the attention mechanism in finding global dependencies, a new class of transformers has emerged. The Detection Transformer (DETR) \cite{zheng2020end} used a ResNet \cite{he2016deep} backbone and a Bipartide matching loss, surpassed previous techniques such as Fast-Mask-R-CNN \cite{girshick2015fast}, which only relies on a region proposal network. DETR was later extended to perform panoptic segmentation with the help of an additional mask head. 
\begin{figure*}[t!]
    \centering
    \includegraphics[width=\textwidth]{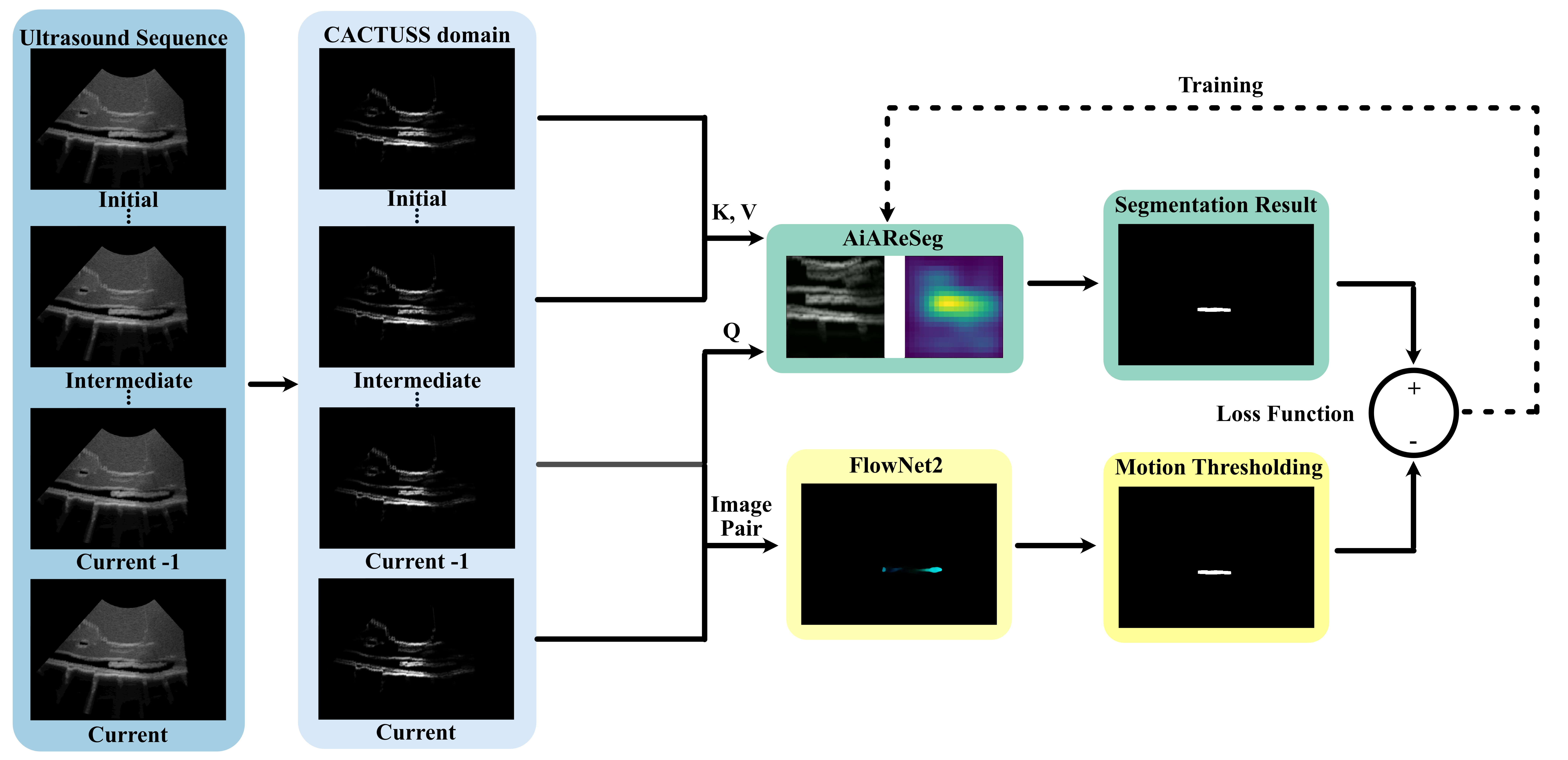}
    \caption{Detailed pipeline of CathFlow. Representative inputs and outputs to each module are shown in the diagram. The K, V and Q, explained in \cite{ranne2023aiareseg}, correspond to the keys, values and queries, respectively. K are feature maps from the initial and intermediate frames, V are the K masked using segmentation results from the previous frame, and Q are features from the current search frame.}
    \label{fig:CathFlow}
\end{figure*}

\subsection{Catheter detection and tracking}

While object detection and tracking in natural images have been extensively studied, applying those techniques to US-guided catheterisation has been limited. Nevertheless, many works chose combined learning and model-based methods to achieve needle detection. Mathiassen et al. \cite{mathiassen2016robust} used Kalman and particle filters to estimate needle axis and tip position, with the help of an initial user input. Chatelain et al. \cite{chatelain2013real} and Zhao et al. \cite{zhao2014biopsy} proposed ROI-RK. They used RANSAC to approximate the position of the needle axis and PCA to refine the initial samples. When it comes to catheter detection, a different approach was adopted by Langsch et al. \cite{langsch2019robotic}, who used a synthetic reference for tracking by template matching. 

 On the other hand, Kim et al. \cite{kim2022learning} tackled the problem of active acoustic catheter (AAC) detection in echocardiography by first using UNet to find the left ventricle, then filtering colors generated by the AAC, before finally performing thresholding to find the catheter. Yang et al. \cite{yang2019catheter} proposed 3D US catheter segmentation using Shared-ConvNet, using CNNs with shared weights across all imaging planes and performing binary classification for each voxel. On the whole, besides our previous work, literature on catheter segmentation in AI-driven US-guided MIES is limited.

\subsection{AiAReSeg and the Attention in Attention Mechanism}

An inherent characteristic of US that has currently been neglected is that it comes in the form of a sequence, where features in a past frame are propagated to the next, or different views of the same feature appear over many frames. In a clinical scenario, this may occur when the clinician is scanning along a pre-planned path or moving the probe back and forth, using information from previous frames to help locate the next ones and interpret the image.

To capture this behaviour, Ranne et al. has introduced the Attention in Attention + ResNet for Segmentation architecture (AiAReSeg) \cite{ranne2023aiareseg} which combined information from across the sequence to infer knowledge on the current frame. The architecture combines feature extraction on a local scale using a CNN-based backbone, a 3-branched transformer that self-attends within the initial, intermediate, and current frame, then cross-attends between them to learn the evolution of each feature across depth or time. For mask reconstruction, features from different positions in the sequence were stacked, then performed a 3D convolution to select the features for generating the mask. Finally, the architecture relies upon the AiA module, which treats the attention map as a new feature map, then performs additional self-attention before multiplying by the value. Practically speaking, this distills distant and noisy features from the attention map. 

\subsection{Catheter Simulations and CACTUSS}

As mentioned in Sect. \ref{Sect: Introduction}, obtaining accurate segmentation of US is challenging, due to lack of labels. Meanwhile, open-source datasets of CT images with pixel-level labels are readily available. Burger et al. ~\cite{us_simulator_rendering} developed a simulator that is capable of generating US-like images by casting rays through tissue maps, acquired from labeled CT or MRI and represent US tissue properties in every label.
Ranne et al. \cite{ranne2023aiareseg} built a large dataset of axial US images from CT tissue label maps for training a catheter tracking network.
In their work, an open-source catheterisation engine--CathSim \cite{jianu2022cathsim} was used to capture the physical behaviour of catheter-tissue interaction, then the catheter positions were mapped back into CT domain, followed by the CT-US simulation. 

Recently, Velikova et al. has demonstrated the use of a novel common anatomical CT-US space (CACTUSS) to improve segmentation performance \cite{velikova2024cactuss}. They generated simulated images in a new intermediate representation domain, used to facilitate training. During deployment, real domain images are translated into the CACTUSS domain, where segmentation is performed. In this way, the domain gap between real and simulated images was minimised and the segmentation performance was optimised. Via visual inspection of the translation results, CACTUSS filters out unwanted domain features, such as artifacts and noise, while also standardising the image intensity distribution and texture.

\subsection{Unsupervised motion segmentation}

Catheterisation is a procedure where the MIES instruments are inserted into vasculature with guidance. Naturally, such instruments move inside of the anatomy, and their motion is captured via imaging. Capturing this motion then harnessing it for segmentation forms the core of our framework.

In sequences of images with continuous motion, optical flow can be calculated from the displacement of each pixel in adjacent images \cite{horn1981determining}. Based on this, Meunier and Bouthemy intoduced an unsupervised UNet-like framework \cite{meunier2023unsupervised} which segments key objects in a video by processesing 3D volumes of optical flow and producing segmentation masks. Backpropagation of the losses was performed in the optical flow space, where the final loss included a flow consistency term, and a regularisation term to ensure temporal consistency.

Another approach proposed by Choudhury et al. \cite{choudhury2022guess}, named Guess What Moves (GWM), leverages optical flow and image features to train a per-pixel segementation network. The architecture consists of two branches, one which predicts an initial guess of optical flow via RAFT \cite{teed2020raft}, while the second is the segmentation backbone. During training, segmentation mask output proposals are provided and clustered, before being compounded into the final prediction. This prediction is used to further define the next flow estimate, producing a reconstruction flow, taking the current motion context into account for the next segmentation. 

\section{METHODOLOGY} \label{Sect: Methodology}

\subsection{Data Acquisition}

\begin{figure}[t!]
    \centering
    \includegraphics[width=0.9\columnwidth]{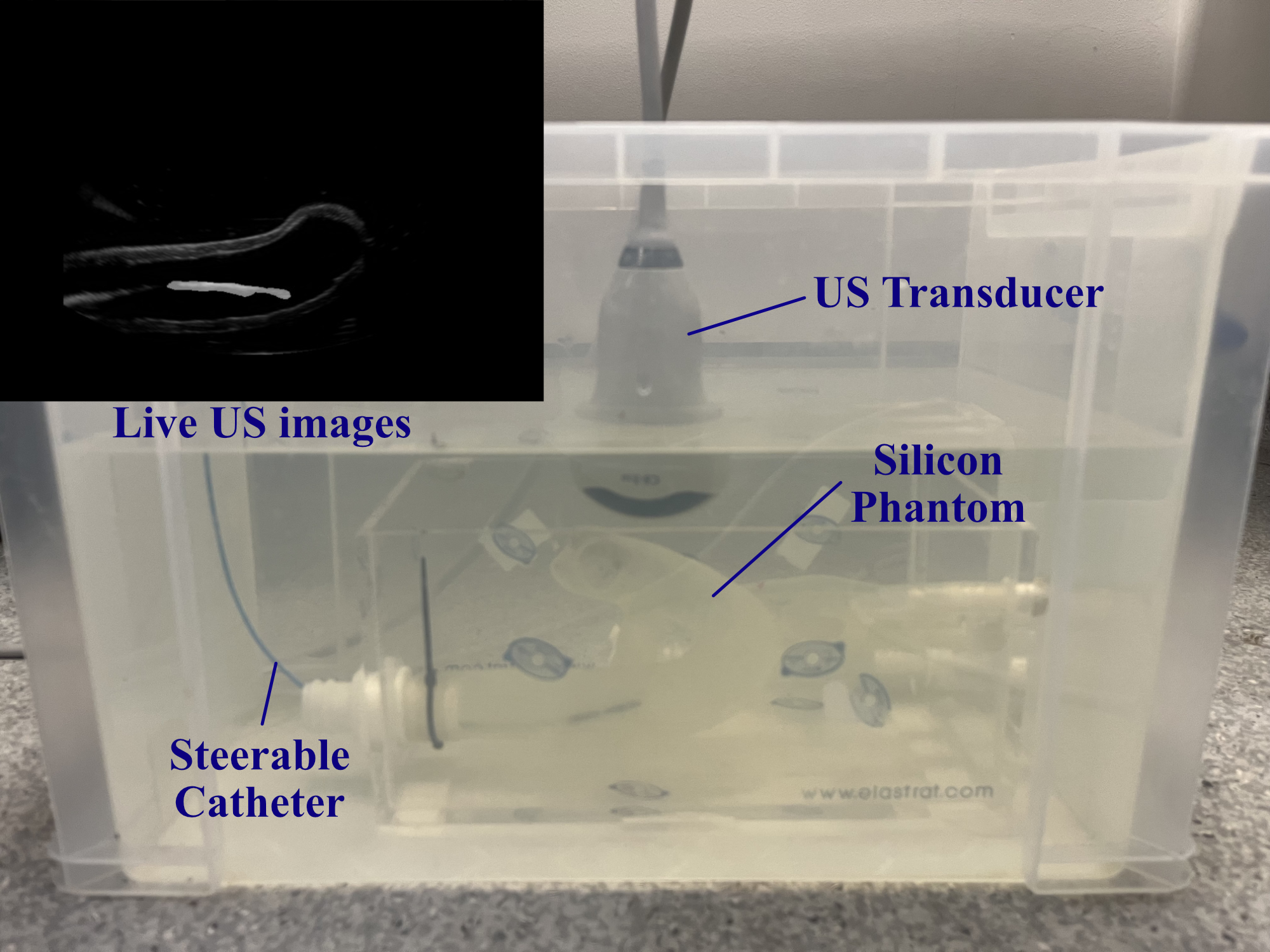}
    \caption{Phantom setup for collecting images}
    \vspace{-5mm}
    \label{fig:Phantom}
\end{figure}

In this work, we used two datasets. The first is a synthetic US dataset, simulated from CT label maps from a public repository\footnote{https://www.synapse.org/\#!Synapse:syn3193805/wiki/89480}. The dataset was generated following the same pipeline as in \cite{ranne2023aiareseg} using ImFusion Suite software\footnote{ImFusion GmbH, Munich, Germany}. Catheterisation was simulated via CathSim \cite{jianu2022cathsim}, where mesh models of the aorta were manually segmented and imported to create the interactive environment. Details of the acquisition process were described in a prior work \cite{ranne2023aiareseg}, with the difference in acquiring longitudinal images. Synthetic US images from 4 CT label maps are used for training, and 1 for evaluation. A second dataset for testing is acquired via a Zonare Z One US machine. These images are from an aortic phantom, placed in a water tank (Fig.\ref{fig:Phantom}). Settings of the US machine were fine-tuned to minimise artifacts and noise, which resemble that of a CACTUSS domain image. Acqusition was done at a depth of 12.0cm, gain setting of G44, and frequency of 7MHz. A representative image acquired via this method is shown alongside the phantom in Fig.~\ref{fig:Phantom}.

\subsection{CathFlow: Self-supervised segmentation} \label{Subsection: CathFlow}

The details of our implemented CathFlow pipeline is shown in Fig.~\ref{fig:CathFlow} and consists of 3 main components.

\emph{Component 1 - CACTUSS (Blue) }: 
In this step, a ray-casting algorithm from ImFusion Suite is used to simulate US images. Input to the algorithm is a label map with 6 acoustic parameters -- speed of sound c, acoustic impedance Z, attenuation coefficient $\alpha$ and speckle distribution parameters, which are needed to mimic the characteristics of tissues in ultrasound. With that said, simulating images with a high degree of similarity to real images does not necessarily aid segmentation performance. Hence, CACTUSS sets tissue-specific speckle parameters to zero, effectively rendering tissues black, and leaving only bright boundaries behind. When used as a pre-processing tool, CACTUSS can translate US images into an easily segmentable domain with much less noise, thereby assisting the proposed segmentation pipeline. 

\emph{Component 2 - FlowNet2 and mask generation (Yellow)}: Since obtaining a large labelled dataset is time-consuming, we utilise motion captured with the optical flow. In this way we automatically generate labels for the images without any external input.
A moving catheter inside of a stationary US image constitutes an opportunity for motion detection algorithms, such as the Farneb{\"a}ck method \cite{farneback2003two}, or networks, such as FlowNet2 \cite{ilg2017flownet}, RAFT \cite{teed2020raft}, and PWC-Net \cite{sun2018pwc} to extract optical flow from the sequence. While a detailed comparison between the aforementioned methods on our dataset can be found in Sect. \ref{Sect: Optical flow}, in this work we have selected FlowNet2 as the best choice due to its ability to extract large and slow-moving objects with large receptive fields, enabled by its CNN-based architecture. 

The key innovation of FlowNet2 is its cascaded and deep architecture, consisting of a FlowNet-C (correlation), two FlowNet-S (simple), a FlowNet-SD (small displacement), and a Fusion block. Both FlowNet-S and FlowNet-SD consist of a standard Fully Convolutional Network (FCN) like architecture, where two adjacent frames are stacked and analysed. Here, progressively smaller kernels were used for feature extraction, and a refinement module built skip connections between the up-scaling decoder with the down-sampling encoder to supply sufficient features for flow reconstruction. On the other hand, FlowNet-C focuses on finding the correlation between two patches on the input images, defined by Eq. \ref{Eq:Correlation}:
\vspace{-1mm}
\begin{equation} \label{Eq:Correlation}
c(\mathrm{\textbf{x}_1},\mathrm{\textbf{x}_2}) = \sum_{\mathbf{o}\epsilon [-k,k]\times [-k,k]}^{} \left< \mathbf{f}_1 (\textbf{x}_1 + \textbf{o}),\textbf{f}_2(\textbf{x}_2 + \textbf{o})\right>
\end{equation}

Where $\textbf{f}_1$ and $\textbf{f}_2$ correspond to the two square image patches with center points $\textbf{x}_1$ and $\textbf{x}_2$, and with the angle brackets corresponding to the convolution procedure performed in neural networks, except that the kernel is swapped for an image patch, and the patch not being trainable. The fusion block adjusts the resolution of the small displacement flow with the flow output from a cascade of FlowNet-CSS to further refine the prediction. Thereafter, a simple motion thresholding is performed to remove background noise with small motion, then the flow field is converted into a segmentation mask via a simple conditional statement, where any point in the flow field with detected motion is labelled with 1. Given an optical flow field \( F \) of size \( (H, W, 2) \), where \( F_{i,j} = (u_{i,j}, v_{i,j}) \) 
represents the flow at pixel \( (i, j) \), and a threshold value \( T \), we define 
a binary segmentation mask \( M \) as follows:

\vspace{-1mm}
\begin{equation} \label{eqn:Threshold}
M_{i,j} =
\begin{cases} 
1 & \text{if } |u_{i,j}| > T \text{ or } |v_{i,j}| > T \\
0 & \text{otherwise}
\end{cases}
\end{equation}

Due to significant background noise, we use different thresholding values for phantom datasets compared to synthetic ones. In the following evaluation and training, if not mentioned, \( T = 0.2\) for synthetic datasets and \( T = 1\) for phantom datasets as default.

\emph{Component 3 - AiAReSeg (Green)}: 
Given an estimation for the segmentation masks, any segmentation pipeline is theoretically applicable. However, in prior works, AiAReSeg stood out as a strong contender, achieving state-of-the-art performance in axial synthetic US images. By leveraging its short-term/long-term cross-attention modules in the transformer decoder, and its 3D deconvolution in the segmentation head, we have demonstrated that the network is effective in retaining temporal or volumetric information during mask prediction, outperforming UNet and thresholding methods that are trained on independent image-mask pairs. Thus, it was chosen to perform segmentation. Since FlowNet2 weights remain frozen, and the only element of the network that is being trained is AiAReSeg, we therefore made no modification to its loss function as introduced in AiAReSeg, consisting of a weighted sum of the binary cross entropy, the dice loss, and the L2 distance.

\subsection{Data Processing}
Prior to segmentation, the synthetic data is first translated to CACTUSS domain to remove intrinsic US noise, which affects the segmentation pipeline. 
FlowNet2 is then deployed on the restructured dataset to generate optical flow. When there are stationary consecutive frames in a sequence, FlowNet2 will generate noisy predictions across the entire frame, while flow from frames with catheter movement would appear to be mostly clean with only significant flow signal from the catheter region. Before training, we filter out stationary frames by abusing the characteristics through thresholding. Meanwhile, we generate a bounding box for the catheter at each individual frame, in order to crop the data, such that previous predictions can help the model better divert its focus and improve the attention results. During training, data augmentations including normalization, cropping, flipping, and rotation, were introduced to boost the model's generalisability. The same augmentation is applied to the search frame, optical flow, and reference frames.

\subsection{Inference}
\begin{figure}[t!]
    \centering
    \includegraphics[width=\columnwidth]{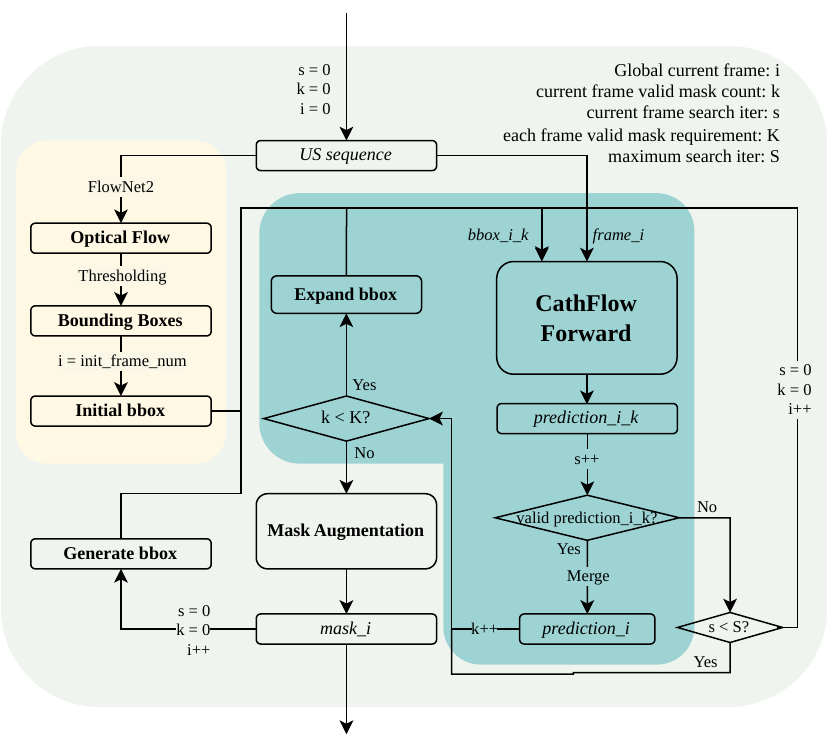}
    \caption{The Inference Pipeline of CathFlow}
    \label{fig:Inference}
\end{figure}

Fig.~\ref{fig:Inference} presents the inferencing pipeline for catheter localisation. The green section indicates the main inference loop, while the yellow section highlights how the initial frame of each inference sequence is determined. Finding the first frame containing a catheter is essential for two reasons. First, we can avoid performing inference on static frames, reducing unnecessary computations. Second, since we use the prediction from the previous frame to generate the bounding box for the current frame. Bounding box quality is essential for better prediction, as it guides the model to focus on regions where the catheter is most likely to be. 

The bounding box is generated using the same aforementioned threshold scheme (Eqn. \ref{eqn:Threshold}), where we directly extract a bounding box for the segmented region. Since FlowNet2 is prone to produce noisy outputs for static frames, these frames can be thresholded by using \( T = 0.2\) for synthetic dataset and \( T = 1\)  for phantom datasets. If the thresholded flow field is empty, there would be no corresponding bounding box, indicating the absence of a catheter. 


Once the initial bounding box is determined, CathFlow’s inference loop is performed (blue section). The forward pass of CathFlow takes the current frame ($i$) and the bounding box of the previous segmentation as input. For each individual frame, we keep track of ($s$): the iteration of inference conducted, and ($k$): the amount of valid predictions. A prediction is ``valid" when a binary prediction has a foreground area larger than 200 pixels. We find it to be a sensible empirical threshold to distinguish a confident catheter prediction, and noisy predictions. During inferencing, the initial bounding box is passed into CathFlow to produce a prediction. If the prediction is deemed valid, we merge it to form prediction$_i$, an aggregate of valid predictions for the current frame. Else, the bounding box expands in size by a factor, which increases as $s$ increases, and seeks to obtain another prediction from a larger scale. This iterative process ensures that when a sudden, large movement of the catheter occurs, our model not only focuses on the area where the previous bounding box was but also tries to acquire the largest prediction possible. Through the strategic expansion of the bounding box and mask augmentation, CathFlow refines its focus, thereby enhancing the localisation accuracy of the catheter in successive frames. 

\section{EXPERIMENTAL EVALUATIONS}\label{Sect: evaluation}
\begin{figure*}[t!]
    \centering
    \includegraphics[width=0.95\textwidth]{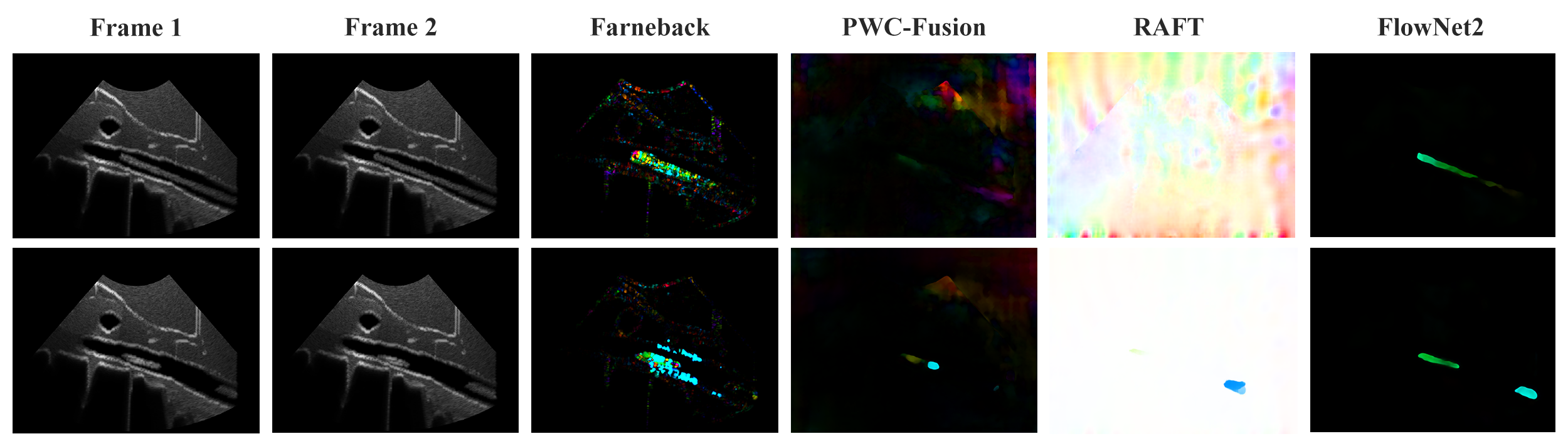}
    \caption{Different optical flow generation methods and respective generated optical flow on ultrasound sequence.}
    \label{fig:OpticalFlow}
\end{figure*}

\subsection{Optical Flow Generation} \label{Sect: Optical flow}

Model-based methods, including Farneb{\"a}ck and deep learning-based methods including PWC-Fusion \cite{ren_fusion_2018} (a variant of PWC-Net), RAFT, and FlowNet2 are explored for optical flow generation. All methods are evaluated on unprocessed US sequences for fairness. In Fig.~\ref{fig:OpticalFlow}, we demonstrate qualitative results of the performance of each method on an US sequence with a stationary probe. Farneb$\ddot{a}$ck shows promise for catheter segmentation, however, the kernel size and parameters heavily impact the quality of the generated flow. Despite extensive fine-tuning, the optical flow remains to be either spotty or overly blurry.
PWC-Net and PWC-Fusion are also studied, however the outputs are less than ideal. PWC-Net has a noticeably long inference time (0.75fps) compared to PWC-Fusion and less detailed outputs. PWC-Fusion is faster but it is still heavily affected by background noise, thus producing less significant flows. The most up to date optical flow prediction architecture we trialed - RAFT, has also performed poorly. Its trait of maintaining a high-resolution flow field is precisely what limits its performance. Due to the noisy nature of US, the sequence usually has large number of small displacements in the background region, leading to an unusable flow in most scenarios. In all of our tests, FlowNet2 consistently delivers the cleanest optical flow and maintains stable segmentation of the catheter. As FlowNet2 creates artificial noise in its prediction if there are no significant catheters movements in the frame, we filter small flow components out via thresholding. This feature is also utilised during inferecing to identify the initial frame where the catheter emerges. 

\subsection{Training details}
Experiments were conducted on a workstation with NVIDIA GeForce RTX 4060Ti 16G, 32G RAM, and AMD Ryzen 5800X.
A total of 5 US scans are synthesized, each scan consists of 100 sequences, and each sequence consists of from 72 to 163 frames, sums up to a total of 66900 frames. 4 of the 5 datasets are used for training (50600 frames), and the remaining one, consisting of 16300 frames for evaluation.

\subsection{Benchmarking details}

Our model is evaluated against other similar unsupervised segmentation models including the nnU-Net \cite{isensee2021nnu} trained in an unsupervised fashion, and the Guess-What-Moves model \cite{choudhury2022guess} with MaskFormer \cite{cheng_per-pixel_2021} backbone. Metrics, including Dice Score and mean average error, are considered. The synthetic benchmark is performed on a dataset generated along with the training set using the aforementioned generation pipeline. The phantom dataset is an accumulation of five individual scans on the same aorta phantom underwater, giving it a total of 3036 frames.

\subsubsection{CathFlow} Our proposed model is trained in an end-to-end fashion. Weights are initialized from a pre-trained AiAReSeg model at 500 epochs. We discard the last 2 layers of the AiA transformer module and all weights in the segmentation head. CathFlow is trained for 100 epochs with a step learning rate scheduler starting at $1\times10^{-3}$. 

\subsubsection{Guess What Moves} We initialised the GWM model with MaskFormer as backbone with the best checkpoint provided by the author trained on DAVIS dataset \footnote{https://davischallenge.org/davis2017/code.html}. This checkpoint is further fine-tuned by the same training set used for training CathFlow for 40000 iterations with a base learning rate of $1\times 10^{-4}$, while using an unfreeze schedule of [(1, 10), (0, 2000), (-1, 5000)]. Since GWM requires pairs of optical flow for training, optical flows of stride -1 are generated using the same FlowNet2 checkpoint for all training and evaluation frames.

\subsubsection{nnU-Net} 

A dynamically configured nnU-Net implemented as part of the MONAI library was used \cite{cardoso2022monai}. In comparison with the standard UNet, nnUNet is able to dynamically adapt its hyperparameters to best fit the task at hand. The nnUNet is trained with a learning rate of $1\times10^{-5}$, initialised with kernel sizes of 7, 5, 3, 3 in its encoder, with a kernel size of 2, 2, 1 in the decoder. Training was done with segmentation mask estimates generated in CathFlow.

\subsubsection{FlowNet2}
In order to demonstrate the efficacy of our model, we also benchmark the thresholded optical flow generated by FlowNet2 for comparison. Here, the optical flow is thresholded by the same principle as during training,  using \( T = 0.2\) for synthetic datasets and \( T = 1\)  for phantom datasets. This means we deem all pixels whose horizontal $u$ and vertical $v$ flow components are smaller than the threshold to be the background and the remaining to be the foreground. 

\begin{table}[t]
    \centering
    
    \caption{Evaluation on synthetic dataset}
    \begin{tabular}{|l|c|c|c|c|}
        \hline
        Metric & Dice Score  & MAE \\
        \hline
        GWM & 58.4$\pm$ 0.213 & 0.1207 $\pm$0.1217 \\
        nnU-Net & 57.6$\pm$ 0.152 & 0.0038 $\pm$ 0.0016 \\
        nnU-Net* & 66.7 $\pm$ 0.119 & 0.0032 $\pm$ 0.0014\\
        FlowNet2 & 67.4 $\pm$ 0.113 & 0.0021 $\pm$ 0.002\\
        \hline
        \textbf{Ours} & \textbf{72.8$\pm$0.199} & \textbf{0.0022$\pm$0.0020} \\
        \hline
    \end{tabular}
    \label{tab:eval-synthetic}
\end{table}

\begin{table}[t]
    \centering
    
    \caption{Evaluation on phantom dataset}
    \begin{tabular}{|l|c|c|c|c|}
        \hline
        Metric & Dice Score  & MAE \\
        \hline
        GWM & 3.7$\pm$ 0.048 & 0.0324 $\pm $0.0525 \\
        nnU-Net & 3.7$\pm$ 0.026 & 0.0102 $\pm$ 0.0020 \\
        nnU-Net* & 2.4$\pm$ 0.026 & 0.0116  $\pm$ 0.0010\\
        FlowNet2 & 14.3 $\pm$ 0.076 & 0.0183 $\pm$ 0.0039\\
        \hline
        \textbf{Ours} & \textbf{41.9$\pm$5.6760} & \textbf{0.0051$\pm$0.0007} \\
        \hline
    \end{tabular}
    \label{tab:eval-phantom}
\end{table}
\section{RESULTS}\label{Sect: results}

In Tab.~\ref{tab:eval-synthetic} and Tab.~\ref{tab:eval-phantom}, we present the segmentation results in the synthetic and phantom datasets, respectively. Evidently, CathFlow with its AiAReSeg backbone decisively outperforms its competitors, averaging to a mean Dice Score of 72.8 for the synthetic set and 41.9 for the phantom set. We presented two cases of nnU-Net in this study. The first which computes the average dice metric for the complete unfiltered dataset, where US sequences may or may not contain a catheter, prompting the model to give a null prediction where there is no catheter. The second (as indicated by asterisks) is for a filtered dataset where every frame contains a catheter. As observed from Tab.~\ref{tab:eval-synthetic}, even without significant advantages, our model still outperformed nnU-Net. Similarly, the mean MAE across both trials also illustrated that CathFlow produces results which are geometrically closer to the ground truth in a pixel-wise manner, where CathFlow produced the lowest MAE of 0.0022 and 0.0051 in synthetic and phantom trials, respectively. Lastly, we also evaluated against segmentation masks directly generated via optical flow, and observed a difference of 5.4 points in synthetic images and 27.6 points in phantom images in favour of CathFlow.

\section{DISCUSSIONS}\label{Sect: discussions}

From the results obtained in Tab.~\ref{tab:eval-synthetic}, we have observed that CathFlow obtained the highest dice score and lowest MAE. This suggests that within the same image domain as the training set - where the images are simulated in the same manner, consisting of the same texture, and preprocessed via CACTUSS - CathFlow is able to outperform its unsupervised counterparts. Our model's main advantage lies in the integration of the AiA module, allowing it to attend better to regions of high weights in an attention map, while ignoring noisy features that may disturb the segmentation results.

When compared with other models, it was clear that the performance improves when temporal information is utilised. First, with GWM, its approach of using a recurrent flow reconstruction as the main training signal to the segmentation network did not provide it with any advantages over our model. Thus, it can be inferred that improving the quality of the flow estimation was not able to aid the main segmentation network, and that the flows generated via FlowNet2 were sufficient to generate a single-shot estimation for our framework. Similar results were observed with the nnU-Net, with both a general unfiltered sequence and a filtered sequence with catheters in every frame. The significant drop in performance was most likely due to nnU-Net assuming that every frame contains a catheter regardless of the features presented to it, thus incorrectly assigning labels. More importantly, we evaluated our model against thresholded FlowNet2, the training signal for our pipeline, and found that it has been surpassed, indicating the significance of AiAReSeg in capturing temporal features, and refining predictions overtime. Overall, both models in-comparison only examine features in one frame for mask generation. however the temporal awareness modules in AiAReSeg, such as the LT/ST cross attention, and 3D deconvolution modules, which is able to infer the changes in the features across time, provide our model the edge. The performance of our model was limited by the splitting of mask segments due to the noisy nature of ultrasound optical flow, which is not present in ground truth masks. Potential future works that investigate techniques to join such disjointed portions together may be explored.

Finally, evaluation against phantom images has highlighted some strengths of our model, albeit also exposing several limitations. Judging by the metrics, our model still outperforms its rivals. In the case where all other models failed to segment, our model was still able to generate better results. However, a score of 41.9 indicates limited generalisation capabilities, in contrast to the supervised AiAReSeg model, likely due to the substantially more difficult task of segmenting in transverse images than the axial case, as well as the quality of the segmentation masks obtained from optical flows. The main obstacle was that in the experiments, the catheter touches the wall, making them indistinguishable and thus causing problems for the segmentation pipelines. In fact, since the scanned phantom was placed in a water tank, its appearance was similar to the aesthetics of a CACTUSS image, thus we did not have to translate it explicitly. In the event that a full clinical dataset is obtained, we can harness the full capability of the CACTUSS pipeline \cite{velikova2024cactuss} to further improve our performance. This pipeline is optimised for clinical data and is capable of surpassing supervised segmentation methods, as demonstrated by the authors.

\section{CONCLUSIONS}\label{Sect: Conclusions}

In this work, we present a self-supervised catheter segmentation approach in iUS-guided endovascular procedures. This framework is built upon the assumption that such image sequences consist of a stationary transducer and a moving catheter advancing through the aorta, and such motion features can be extracted as optical flows, then converted into segmentation mask estimates. We continued our previous line-of-work in synthesising catheterisation US sequences using a CT-to-US ray-casting and physics-based simulator. Then, we translated our images into the CACTUSS domain, filtering out unwanted features and easing the task. Following a qualitative comparison of the quality of the flow reconstruction, FlowNet2 was selected, as it strikes a satisfactory balance between the cleanliness of the output, and the inference speed. Using this method, we trained an improved version of AiAReSeg, and finally benchmarked its performance in both synthetic and phantom images, against state-of-the-art unsupervised frameworks, and the nnU-Net, trained in the same self-supervised manner. Our results highlighted the feasibility of the framework to translate from sim-to-real, outperforming its rivals by a substantial margin. This work presents an important milestone towards automatic labelling and segmentation in a surgical workflow, and has the potential to be integrated into MIES following fine-tuning and validation on clinical data.






\bibliographystyle{unsrt}
\bibliography{refs}

\begin{thebibliography}{10}

\bibitem{kaplan2016kaplan}
Joel~A Kaplan.
\newblock {\em Kaplan's cardiac anesthesia: In cardiac and noncardiac surgery}.
\newblock Elsevier Health Sciences, 2016.

\bibitem{ullery2018epidemiology}
Brant~W Ullery, Richard~L Hallett, and Dominik Fleischmann.
\newblock Epidemiology and contemporary management of abdominal aortic aneurysms.
\newblock {\em Abdominal Radiology}, 43:1032--1043, 2018.

\bibitem{weisbord2005radiocontrast}
Steven~D Weisbord and Paul~M Palevsky.
\newblock Radiocontrast-induced acute renal failure.
\newblock {\em Journal of Intensive Care Medicine}, 20(2):63--75, 2005.

\bibitem{kopp2010first}
Reinhard Kopp, Werner Z{\"u}rn, Rolf Weidenhagen, Georgios Meimarakis, and Dirk~A Clevert.
\newblock First experience using intraoperative contrast-enhanced ultrasound during endovascular aneurysm repair for infrarenal aortic aneurysms.
\newblock {\em Journal of vascular surgery}, 51(5):1103--1110, 2010.

\bibitem{wakabayashi2013ultrasound}
Masanori Wakabayashi, Sayaka Hanada, Hiroyuki Nakano, and Tsunemichi Wakabayashi.
\newblock Ultrasound-guided endovascular treatment for vascular access malfunction: results in 4896 cases.
\newblock {\em The Journal of Vascular Access}, 14(3):225--230, 2013.

\bibitem{gibbs2011ultrasound}
Vivien Gibbs, David Cole, and Antonio Sassano.
\newblock {\em Ultrasound physics and technology: how, why and when}.
\newblock Elsevier Health Sciences, 2011.

\bibitem{ranne2023aiareseg}
Alex Ranne, Yordanka Velikova, Nassir Navab, and Ferdinando Rodriguez~y Baena.
\newblock Aiareseg: Catheter detection and segmentation in interventional ultrasound using transformers.
\newblock In {\em 2024 IEEE International Conference on Robotics and Automation (ICRA)}, 2024.

\bibitem{velikova2024cactuss}
Yordanka Velikova, Walter Simson, Mohammad~Farid Azampour, Philipp Paprottka, and Nassir Navab.
\newblock Cactuss: Common anatomical ct-us space for us examinations.
\newblock {\em International Journal of Computer Assisted Radiology and Surgery}, pages 1--9, 2024.

\bibitem{zaffino2019fully}
Paolo Zaffino, Guillaume Pernelle, Andre Mastmeyer, Alireza Mehrtash, Hongtao Zhang, Ron Kikinis, Tina Kapur, and Maria~Francesca Spadea.
\newblock Fully automatic catheter segmentation in mri with 3d convolutional neural networks: application to mri-guided gynecologic brachytherapy.
\newblock {\em Physics in Medicine \& Biology}, 64(16):165008, 2019.

\bibitem{nguyen2020end}
Anh Nguyen, Dennis Kundrat, Giulio Dagnino, Wenqiang Chi, Mohamed~EMK Abdelaziz, Yao Guo, YingLiang Ma, Trevor~MY Kwok, Celia Riga, and Guang-Zhong Yang.
\newblock End-to-end real-time catheter segmentation with optical flow-guided warping during endovascular intervention.
\newblock In {\em 2020 IEEE International Conference on Robotics and Automation (ICRA)}, pages 9967--9973. IEEE, 2020.

\bibitem{litjens2017survey}
Geert Litjens, Thijs Kooi, Babak~Ehteshami Bejnordi, Arnaud Arindra~Adiyoso Setio, Francesco Ciompi, Mohsen Ghafoorian, Jeroen~Awm Van Der~Laak, Bram Van~Ginneken, and Clara~I S{\'a}nchez.
\newblock A survey on deep learning in medical image analysis.
\newblock {\em Medical image analysis}, 42:60--88, 2017.

\bibitem{jiang2021autonomous}
Zhongliang Jiang, Zhenyu Li, Matthias Grimm, Mingchuan Zhou, Marco Esposito, Wolfgang Wein, Walter Stechele, Thomas Wendler, and Nassir Navab.
\newblock Autonomous robotic screening of tubular structures based only on real-time ultrasound imaging feedback.
\newblock {\em IEEE Transactions on Industrial Electronics}, 69(7):7064--7075, 2021.

\bibitem{mishra2018ultrasound}
Deepak Mishra, Santanu Chaudhury, Mukul Sarkar, and Arvinder~Singh Soin.
\newblock Ultrasound image segmentation: a deeply supervised network with attention to boundaries.
\newblock {\em IEEE Transactions on Biomedical Engineering}, 66(6):1637--1648, 2018.

\bibitem{ronneberger2015u}
Olaf Ronneberger, Philipp Fischer, and Thomas Brox.
\newblock U-net: Convolutional networks for biomedical image segmentation.
\newblock In {\em Medical Image Computing and Computer-Assisted Intervention-MICCAI}, 2015.

\bibitem{isensee2021nnu}
Fabian Isensee, Paul~F Jaeger, Simon~AA Kohl, Jens Petersen, and Klaus~H Maier-Hein.
\newblock nnu-net: a self-configuring method for deep learning-based biomedical image segmentation.
\newblock {\em Nature methods}, 18(2):203--211, 2021.

\bibitem{chorowski2015attention}
Jan~K Chorowski, Dzmitry Bahdanau, Dmitriy Serdyuk, Kyunghyun Cho, and Yoshua Bengio.
\newblock Attention-based models for speech recognition.
\newblock {\em Advances in neural information processing systems}, 28, 2015.

\bibitem{vaswani2017attention}
Ashish Vaswani, Noam Shazeer, Niki Parmar, Jakob Uszkoreit, Llion Jones, Aidan~N Gomez, {\L}ukasz Kaiser, and Illia Polosukhin.
\newblock Attention is all you need.
\newblock {\em Advances in neural information processing systems}, 30, 2017.

\bibitem{dosovitskiy2020vit}
Alexey Dosovitskiy, Lucas Beyer, Alexander Kolesnikov, Dirk Weissenborn, Xiaohua Zhai, Thomas Unterthiner, Mostafa Dehghani, Matthias Minderer, Georg Heigold, Sylvain Gelly, Jakob Uszkoreit, and Neil Houlsby.
\newblock An image is worth 16x16 words: Transformers for image recognition at scale.
\newblock {\em ICLR}, 2021.

\bibitem{zheng2021rethinking}
Sixiao Zheng, Jiachen Lu, Hengshuang Zhao, Xiatian Zhu, Zekun Luo, Yabiao Wang, Yanwei Fu, Jianfeng Feng, Tao Xiang, Philip~HS Torr, et~al.
\newblock Rethinking semantic segmentation from a sequence-to-sequence perspective with transformers.
\newblock In {\em Proceedings of the IEEE/CVF conference on computer vision and pattern recognition}, 2021.

\bibitem{zheng2020end}
Minghang Zheng, Peng Gao, Renrui Zhang, Xiaogang Wang, Hongsheng Li, and Hao Dong.
\newblock End-to-end object detection with adaptive clustering transformer.
\newblock In {\em British Machine Vision Conference}, 2020.

\bibitem{he2016deep}
Kaiming He, Xiangyu Zhang, Shaoqing Ren, and Jian Sun.
\newblock Deep residual learning for image recognition.
\newblock In {\em Proceedings of the IEEE conference on computer vision and pattern recognition}, 2016.

\bibitem{girshick2015fast}
Ross Girshick.
\newblock Fast r-cnn.
\newblock In {\em Proceedings of the IEEE international conference on computer vision}, pages 1440--1448, 2015.

\bibitem{mathiassen2016robust}
Kim Mathiassen, Diego Dall’Alba, Riccardo Muradore, Paolo Fiorini, and Ole~Jakob Elle.
\newblock Robust real-time needle tracking in 2-d ultrasound images using statistical filtering.
\newblock {\em IEEE Transactions on Control Systems Technology}, 25(3):966--978, 2016.

\bibitem{chatelain2013real}
Pierre Chatelain, Alexandre Krupa, and Maud Marchal.
\newblock Real-time needle detection and tracking using a visually servoed 3d ultrasound probe.
\newblock In {\em 2013 IEEE International Conference on Robotics and Automation}, pages 1676--1681. IEEE, 2013.

\bibitem{zhao2014biopsy}
Yue Zhao, Adeline Bernard, Christian Cachard, and Herv{\'e} Liebgott.
\newblock Biopsy needle localization and tracking using roi-rk method.
\newblock In {\em Abstract and Applied Analysis}. Wiley Online Library, 2014.

\bibitem{langsch2019robotic}
Fernanda Langsch, Salvatore Virga, Javier Esteban, R{\"u}diger G{\"o}bl, and Nassir Navab.
\newblock Robotic ultrasound for catheter navigation in endovascular procedures.
\newblock In {\em IEEE/RSJ International Conference on Intelligent Robots and Systems (IROS)}, 2019.

\bibitem{kim2022learning}
Taeouk Kim, Mohammadali Hedayat, Veronica~V Vaitkus, Marek Belohlavek, Vinayak Krishnamurthy, and Iman Borazjani.
\newblock A learning-based, region of interest-tracking algorithm for catheter detection in echocardiography.
\newblock {\em Computerized Medical Imaging \& Graphics}, 2022.

\bibitem{yang2019catheter}
Hongxu Yang, Caifeng Shan, Alexander~F Kolen, and Peter~HN de~With.
\newblock Catheter localization in 3d ultrasound using voxel-of-interest-based convnets for cardiac intervention.
\newblock {\em International journal of computer assisted radiology and surgery}, 14:1069--1077, 2019.

\bibitem{us_simulator_rendering}
Benny Burger, Sascha Bettinghausen, Matthias Radle, and Jürgen Hesser.
\newblock Real-time gpu-based ultrasound simulation using deformable mesh models.
\newblock {\em IEEE Transactions on Medical Imaging}, 2013.

\bibitem{jianu2022cathsim}
Tudor Jianu, Baoru Huang, Mohamed~EMK Abdelaziz, Minh~Nhat Vu, Sebastiano Fichera, Chun-Yi Lee, Pierre Berthet-Rayne, Anh Nguyen, et~al.
\newblock Cathsim: An open-source simulator for autonomous cannulation.
\newblock {\em arXiv preprint arXiv:2208.01455}, 2022.

\bibitem{horn1981determining}
Berthold~KP Horn and Brian~G Schunck.
\newblock Determining optical flow.
\newblock {\em Artificial intelligence}, 17(1-3):185--203, 1981.

\bibitem{meunier2023unsupervised}
Etienne Meunier and Patrick Bouthemy.
\newblock Unsupervised space-time network for temporally-consistent segmentation of multiple motions.
\newblock In {\em Conference on Computer Vision and Pattern Recognition}, 2023.

\bibitem{choudhury2022guess}
Subhabrata Choudhury, Laurynas Karazija, Iro Laina, Andrea Vedaldi, and C.~Rupprecht.
\newblock Guess what moves: Unsupervised video and image segmentation by anticipating motion.
\newblock In {\em British Machine Vision Conference}, 2022.

\bibitem{teed2020raft}
Zachary Teed and Jia Deng.
\newblock Raft: Recurrent all-pairs field transforms for optical flow.
\newblock In {\em European Conference on Computer Vision}, 2020.

\bibitem{farneback2003two}
Gunnar Farneb{\"a}ck.
\newblock Two-frame motion estimation based on polynomial expansion.
\newblock In {\em Image Analysis: 13th Scandinavian Conference}, 2003.

\bibitem{ilg2017flownet}
Eddy Ilg, Nikolaus Mayer, Tonmoy Saikia, Margret Keuper, Alexey Dosovitskiy, and Thomas Brox.
\newblock Flownet 2.0: Evolution of optical flow estimation with deep networks.
\newblock In {\em Proceedings of the IEEE conference on computer vision and pattern recognition}, pages 2462--2470, 2017.

\bibitem{sun2018pwc}
Deqing Sun, Xiaodong Yang, Ming-Yu Liu, and Jan Kautz.
\newblock Pwc-net: Cnns for optical flow using pyramid, warping, and cost volume.
\newblock In {\em IEEE conference on computer vision and pattern recognition}, 2018.

\bibitem{ren_fusion_2018}
Zhile Ren, Orazio Gallo, Deqing Sun, Ming-Hsuan Yang, Erik~B. Sudderth, and Jan Kautz.
\newblock A fusion approach for multi-frame optical flow estimation.
\newblock {\em 2019 IEEE Winter Conference on Applications of Computer Vision (WACV)}.

\bibitem{cheng_per-pixel_2021}
Bowen Cheng, Alexander~G. Schwing, and Alexander Kirillov.
\newblock Per-pixel classification is not all you need for semantic segmentation.
\newblock In {\em Neural Information Processing Systems}, 2021.

\bibitem{cardoso2022monai}
M~Jorge Cardoso, Wenqi Li, Richard Brown, Nic Ma, Eric Kerfoot, Yiheng Wang, Benjamin Murrey, Andriy Myronenko, Can Zhao, Dong Yang, et~al.
\newblock Monai: An open-source framework for deep learning in healthcare.
\newblock {\em arXiv preprint arXiv:2211.02701}, 2022.

\end{thebibliography}
\vfill
\newpage

\end{document}